\documentclass[aps,twocolumn,showpacs,preprintnumbers,amsmath,amssymb,prl]{revtex4-1}

\usepackage[colorlinks=true,bookmarks=false,citecolor=blue,urlcolor=blue]{hyperref} 

\usepackage{amsmath,amssymb}
\usepackage{graphicx}
\usepackage{verbatim} 
\usepackage{color}

\usepackage[english]{babel}
\usepackage{bm}
\usepackage{natbib}
\usepackage{graphicx}

\def\dfrac{\displaystyle\frac}  

\newcommand{\eps}{\varepsilon}

\begin{document}
\title{Trapping and guiding surface plasmons in curved graphene landscapes}

\author{Daria Smirnova$^{1}$}
\author{S. Hossein Mousavi$^{2}$}
\author{Zheng Wang$^{2}$}
\author{Yuri S. Kivshar$^{1}$}
\author{Alexander B. Khanikaev$^{3,4}$}\email[]{akhanikaev@qc.cuny.edu}

\affiliation{
$^{1}$Nonlinear Physics Center, Australian National University, Canberra ACT 0200, Australia\\
$^{2}$Microelectronics Research Center, Cockrell School of Engineering, University of Texas at Austin, Austin, TX 78758 USA\\
$^{3}$Department of Physics, Queens College of The City University of New York, Queens, NY 11367\\
$^{4}$The Graduate Center of The City University of New York, New York, NY 10016}


\begin{abstract}
We demonstrate that graphene placed on top of structured substrates offers a novel approach for trapping and guiding surface plasmons. A monolayer graphene with a spatially varying curvature exhibits an effective trapping potential for graphene plasmons near curved areas such as bumps, humps and wells. We derive the governing equation for describing such localized channel plasmons guided by curved graphene and validate our theory by the first-principle numerical simulations. The proposed confinement mechanism enables plasmon guiding by the regions of maximal curvature, and it offers a versatile platform for manipulating light in planar landscapes. In addition, isolated deformations of graphene such as bumps are shown to support localized surface modes and resonances suggesting a new way to engineer plasmonic metasurfaces.
\end{abstract}

\maketitle

{\em Introduction.} Unprecedented optical properties make graphene a promising plasmonic material with great potential for practical applications ranging from optical routing~\cite{RevGrigorenko,Lozovik_Usp_2012,Review_Jablan,Review_Luo,Abajo_Review_2014} to nonlinear optics~\cite{Glazov}, optomechanics~\cite{optomechanics} and sensing~\cite{biosensing}. In addition to relatively strong and tunable electromagnetic response of this one-atom-thick material, it was shown that at infra-red frequencies graphene supports surface plasmon-polaritons (SPPs) with
extreme confinement and propagation characteristics controllable by doping~\cite{Jablan_2009}. In particular, thanks to the strong dependence of its optical characteristics from electric bias graphene is on the path to be widely used for electro-optical modulation~\cite{RevGrigorenko,modulator}. Possibility of such electrostatic doping enables control over the electrons density at the Fermi level which defines high frequency optical response of graphene~\cite{Wunsch,sptdep,Chen_2012,Fei_2012,Science_2014}. With tremendous success in graphene electronics witnessed recently, these unique capabilities enable integration of plasmonic and electronic devices on the same substrate~\cite{Review_Bao, Mousavi_tuning, Low_Review, Alu_1, Alu_2} and even a design of tunable cloaking devices~\cite{Alu_3}. 

Presently, two of the most common approaches to utilize unique plasmonic response of graphene for guiding applications rely on (i) inhomogeneous gating of graphene~\cite{Engheta, Engheta_PRB_2012} affecting its local plasmonic response, and (ii) subwavelength-scale patterning of graphene layer giving rise to localized plasmonic resonances~\cite{Abajo_Review_2014} .

\begin{figure}[b]
\centering\includegraphics[width=0.98\linewidth]{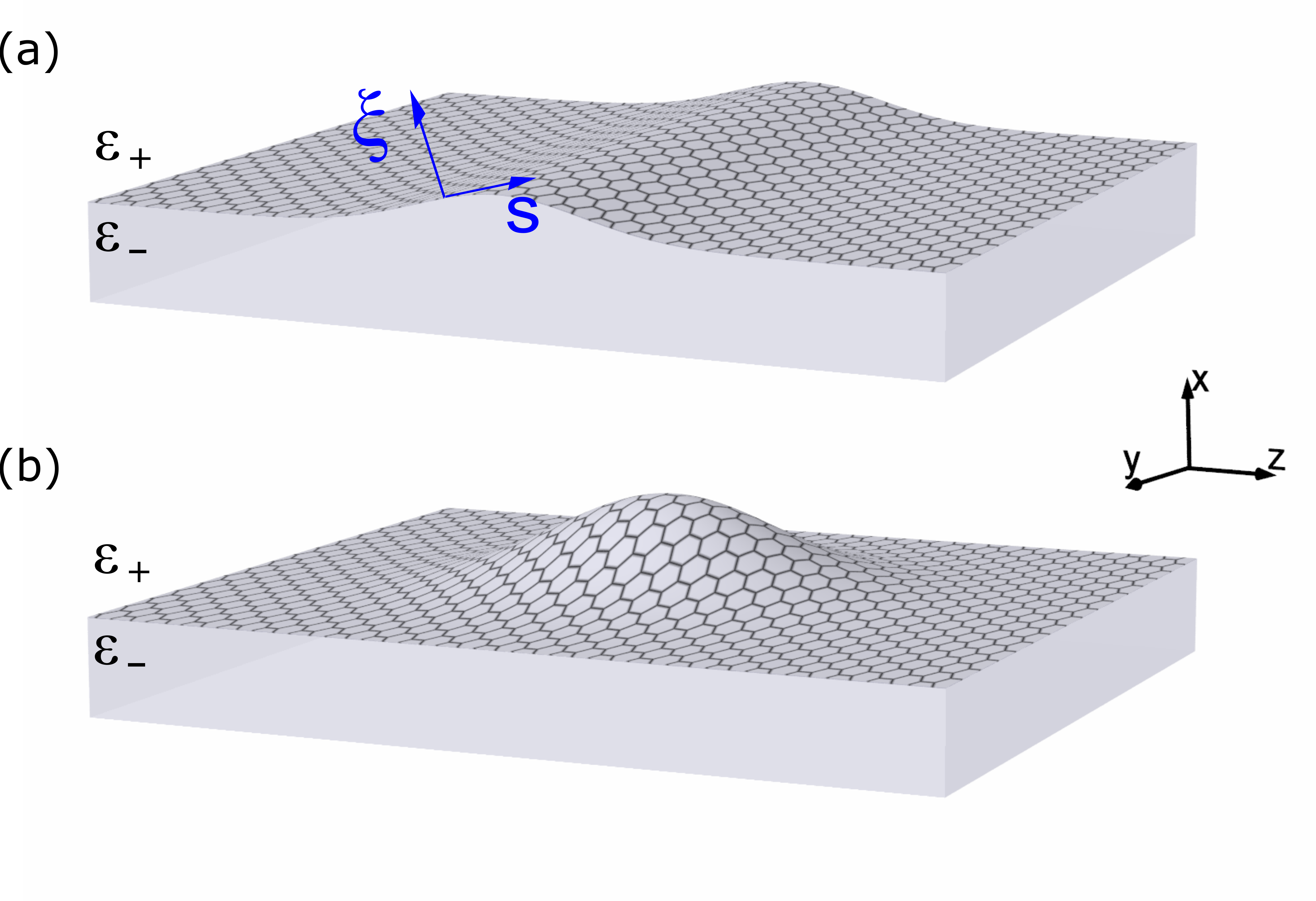}
\caption{(Color online) Schematic view of convex curved substrates covered by graphene and associated global Cartesian and local curvilinear coordinate systems. (a) and (b) correspond to the cases of one- and two-dimensional curved surfaces, respectively.
}
\label{fig:fig1}
\end{figure}

In this Letter, we propose an alternative approach to engineer plasmonic guiding properties of graphene by employing its property to confine and scatter light by curved regions such as extended one-dimensional humps and local two-dimensional deformations. Technologically, such curved graphene profiles, shown schematically in Figs.~\ref{fig:fig1}(a,b), can be fabricated by either transferring graphene onto curved dielectric substrates or grown directly by laser ablation on structured surface of SiC demonstrated recently~\cite{Gerhard_APL_2012}. In particular, we  demonstrate plasmon guiding by a bump of monolayer graphene which effectively operates as a plasmonic channel waveguide~\cite{Maradudin_PRB_1990,Gramotnev_APL_2004,Bozhevolnyi_PRL_2005,Volkov_NL_2007,Moreno_PRL_2008,Bozhevolnyi_OL_2009,Dintinger_OE_2009,Gramotnev_NP_2010,Lee_OE_2011}. In addition, we show that a local isolated deformation of graphene in the form of a bump supports local surface plasmonic resonances. 

It has been already established that a curved graphene surface can play a role of an effective potential for electrons and phonons, and it may support spatially localized modes~\cite{Savin_APL_2009}.  Similarly, we may expect that graphene plasmons can be modified by effective geometric potentials, as this was shown earlier for curved metal-dielectric interfaces~\cite{Longhi_2010}. We notice that the study of the propagation of light in curved space has attracted some special attention due to novel opportunities to guide and focus light~\cite{Peschel_PRA_2008,Peschel_PRL_2010,Peschel_PRA_2010}. Here, we extend this knowledge to the case of plasmons in curved graphene and demonstrate novel approaches to engineer plasmonic
response of curved graphene.

{\em Analytical approach.} As the first step, we explore the capability of trapping light with the use of an analytical approach assuming graphene surface of small one-dimensional curvature. Specifically, we consider a graphene monolayer placed on a curved interface between two dielectric media with permittivities $\eps_{-}$ and $\eps_{+}$, as shown in Fig.~\ref{fig:fig1}(a). Such a graphene-coated surface forms a channel waveguide for surface plasmons, thus enabling a two-dimensional (2D) spatial localization of light, alternative to waveguides created by graphene conductivity modulation~\cite{Engheta, Engheta_PRB_2012}. We base our analytical consideration on Maxwell's equations solved in the adiabatic approximation within the small-wavelength limit.

\begin{figure}[t!]
\centering\includegraphics[width=0.85\linewidth]{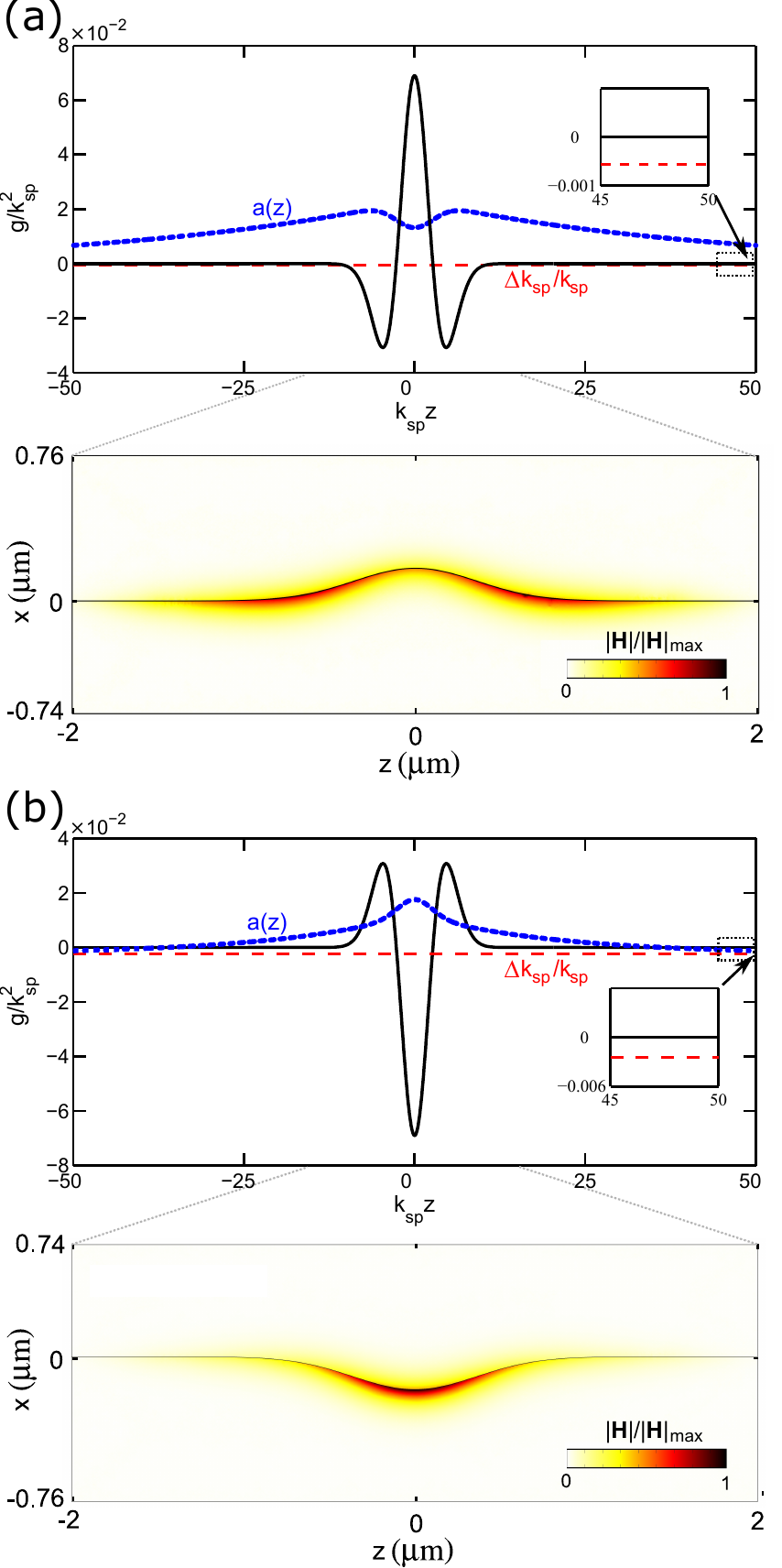}
\caption{(Color online) Top panels in (a,b): Effective potential (black solid curves) and transverse profiles $a(z)$ (a.u.) of the trapped modes (blue dotted curves) calculated for the Gaussian profiles $f(z)=\pm f_0\exp{(-z^2/w^2)}$, $f_0=200$ nm, $w=506$ nm, and dielectric permittivities $\eps_{+} = 1$, $\eps_{-} = 2$ at the frequency $\omega_0/{2\pi} = 16 $ THz. The red dashed line levels the relative correction $\Delta k_{\text{sp}}/k_{\text{sp}}(\omega_0)$ to the wavenumber $k_{\text{sp}}(\omega_0)= 22.4 k_0$ corresponding to the fundamental channel mode. Bottom panels in (a,b): Normalized magnetic field distribution $|{\bf H}|/|{\bf H}|_\text{max}$ of the modes shown in top pannels and obtained with the use of first-principle FEM solver.
}
\label{fig:fig2}
\end{figure}

We assume that an inhomogeneity in the graphene profile is described by a smooth function $x=f(z)$, and we look for modes propagating along the $y$ axis. For convenience,  the curvilinear orthogonal coordinate system $\left\{ \xi , y, s \right\}$ bound to the reference curve $x=f(z)$ is used in place of the Cartesian coordinate system $\left\{ x, y, z \right\}$.
Accepting harmonic $\exp (-i\omega t)$ time-dependence, Maxwell's equations can be written as follows
\begin{equation}
\left \{
\begin{aligned} \label{eq:eqMaxw}
\nabla \times {\bf E}&=ik_{0} {\bf H}\:,\\
\nabla \times {\bf H}&=-ik_{0} \varepsilon (\xi) {\bf E}+\frac{4\pi }{c} \delta (\xi){\sigma } (\omega){\bf E}_{\tau }\:,\\
\end{aligned}
\right.
\end{equation}
where $k_0 = \omega/c$ is the wavenumber in free space, ${\sigma} (\omega)\equiv i \sigma^{(I)}(\omega) +  \sigma^{(R)}(\omega)$ is the linear frequency-dependent surface conductivity of graphene, the function $\varepsilon (\xi)$ describes the dielectric permittivity distribution,
\begin{equation}
\varepsilon(\xi)=
 \begin{cases}
 \varepsilon_{-}\:,&\text{} \xi<0,\\
  \varepsilon_{+}\:,&\text{} \xi>0.
 \end{cases}\\
\end{equation}

The Dirac delta-function $\delta (\xi)$ in Eq.~(\ref{eq:eqMaxw}) indicates that the graphene layer is placed at $\xi=0$, and the subscript $\tau$ refers to the field component tangential to the surface. Equation~(\ref{eq:eqMaxw}) leads to
\begin{equation}
\label{eq:eqWEH}
-\nabla \times \nabla \times {\bf H}+k_{0} ^{2} \varepsilon (\xi){\bf H}=ik_{0}
 \nabla \varepsilon (\xi)\times {\bf E}_{\tau } -
\frac{4\pi }{c} {\sigma} (\omega) \nabla \times \delta (\xi){\bf E}_{\tau }\:,
\end{equation}
where ${\bf E}_{\tau}=\dfrac{i}{k_{0}\varepsilon(\xi)} \left(  \nabla \times {\bf H}\right)_{\tau}$ is continuous at $\xi = 0$, and the vector operations can be calculated using the Lame coefficients given as $h_{\xi}=1$, $h_{y}=1$, $h_s=1-\varkappa(s)\xi \equiv 1 - \xi/R$, where $\varkappa(s) \approx {f''}(z)$ is a signed curvature of the reference cylinder line $x=f(z)$, $R=1/\varkappa(s)$ is the local radius of curvature.

Following the approach employed earlier in Refs.~\cite{Gorbach2013, Coupler_PRB, DissipSoliton_LPR}, we derive the equation for the slowly varying plasmon amplitude, employing the asymptotic description often used in the paraxial optics and physics of optical solitons~\cite{book}.
To develop the consistent perturbation theory, we introduce a small parameter, 
\begin{equation*}
\mu^2 = \text{max}  \left\{  \dfrac{1}{k_0 \left|R\right|}, \left| \dfrac{\sigma^{(R)}}{\sigma^{(I)}}\right| \right\},
\end{equation*}
assuming the losses in graphene and spatial inhomogeneity to be small.
Accordingly, the solution of Eq.~(\ref{eq:eqWEH}) is sought in the form
%
\begin{equation} \label{eq:ansatz}
\begin{aligned}
H_{s } & = \biggl[ \mathsf{ A }(\mu^2 \mathsf y, \mu \mathsf s)h(\xi)
+ \mu^2 \mathsf H^{(2)}(\mu^2 \mathsf y,\mu \mathsf s, \xi) + \ldots \biggr] e^{ik_{\text{sp}} y}\\
& = \biggl[ \mathcal{ A }(s,y)h(\xi)+ H^{(2)}(s,y,\xi) + \ldots \biggr]e^{ik_{\text{sp}}y}\:, \\
H_{y} & = \biggl[ \mu \mathsf{ H }^{(1)}(\mu^2 \mathsf y,\mu \mathsf s, \xi)
+ \mu^3 \mathsf H^{(3)}(\mu^2 \mathsf y,\mu \mathsf s, \xi) + \ldots \biggr]e^{ik_{\text{sp}} y}\\
& =\biggl[ { H }^{(1)}(s,y,\xi)+ H^{(3)}(s,y,\xi) + \ldots\biggr] e^{ik_{\text{sp}} y}\:.
\end{aligned}
\end{equation}
where $\mathcal{ A }(s,y)$ is slowly varying amplitude of the plasmonic mode propagating along the curved channel and $k_{\text{sp}}(\omega)$ satisfies the dispersion relation
\begin{equation} \label{eq:dispflat}
\dfrac{\varepsilon_{+}} {\kappa_{+}} + \dfrac{\varepsilon_{-}} {\kappa_{-}}=-i\dfrac{4\pi}{\omega} \sigma(\omega) =  \dfrac{4\pi}{\omega} \sigma^{(I)} (\omega)\:,
\end{equation}
where $\kappa _{+,-} =\left(k_{\text{sp}}^{2} -k_{0}^2\varepsilon _{+,-}\right)^{1/2}\:$.

Substituting the expression~(\ref{eq:ansatz}) into Eq.~(\ref{eq:eqWEH}) in the zero order in $\mu$ returns the dispersion relation~(\ref{eq:dispflat}) and transverse profile for the plasmon 
on a locally flat graphene,
\begin{equation} \label{eq:H0}
h(\xi) = -ik_0
\begin{cases}
 \dfrac{\varepsilon_{+}} {\kappa_{+}}e^{-\kappa_{+} \xi} ,&\text{ \ $\xi>0$\:,}\\
 -\dfrac{\varepsilon_{-}} {\kappa_{-}}e^{\kappa_{-} \xi} ,&\text{ \ $\xi<0$\:.}
 \end{cases}
\end{equation}
The correction $H^{(1)}$ of the first order in $\mu$ is determined from
$\nabla \cdot {\bf H}= 0$ as
\begin{equation*}
H^{(1)}  = \dfrac{i}{k_{\text{sp}} }\dfrac{\partial \mathcal A}{\partial s}h(\xi)\:.\\
\end{equation*}

In the order of $\mu^2$ the correction $H^{(2)}$ at $\xi \neq 0$ satisfies the equation
\begin{equation} \label{eq:H2}
\begin{aligned}
& \dfrac{d^2 H^{(2)}}{d\xi^2} +  (k_0^2\varepsilon(\xi) - k_{\text{sp}}^2)H^{(2)} = F\:,\\
& F = -\left( 2i k_{\text{sp}}  \dfrac{\partial \mathcal A}{\partial y} + \dfrac{\partial^2 \mathcal A}{\partial s^2}\right) h(\xi)
- \varkappa(s) \mathcal A  \dfrac{\partial h}{\partial \xi} \:.
\end{aligned}
\end{equation}
Matching solutions of Eq.~(\ref{eq:H2}) at the interface $\xi=0$ by using the field boundary conditions
\begin{equation*} \nonumber 
\begin{aligned}
& E^{(2)}_y (\xi = +0 ) = E^{(2)}_y (\xi = -0 ) \equiv E^{(2)}_y ,    \\
& H^{(2)} \! (\xi  \! =  \! +0) \! - \! H^{(2)} \! (\xi \!  =  \! -0 ) \!= \!- \dfrac{4 \pi} {c} \!\left(\! \sigma^{(R)} \! (\omega) \! \mathcal A \!  + \!  i \sigma^{(I)} \!(\omega)\!  E^{(2)}_y \!\right) \!,   %
\end{aligned}
\end{equation*}
where
\begin{equation*}
E^{(2)}_y (\xi = \pm 0 ) = \dfrac{1}{i k_0 \eps_{\pm}} \dfrac{ \partial H^{(2)} }{\partial \xi} (\xi = \pm 0)- \dfrac{1}{R \kappa_{\pm} } \mathcal A \:,
\end{equation*}
finally we find that the amplitude $\mathcal A$ of the plasmon satisfies equation
\begin{equation} \label{eq:eqA}
{2ik_{\text{sp}}\left(\dfrac{\partial \mathcal A}{\partial y} + \gamma \mathcal A \right) +\dfrac{\partial^2 \mathcal A}{\partial s^2} - g(s)\mathcal A = 0\:,}
\end{equation}
where the damping coefficient $\gamma$ is expressed as
\begin{equation*}
\gamma = \dfrac{ \dfrac {4 \pi} {\omega} \sigma^{(R)} (\omega) }
{ \left( \dfrac{\varepsilon_{-}}{\kappa_{-}^3} + \dfrac{\varepsilon_{+}}{\kappa_{+}^3} \right) k_{\text{sp}} } \:,
\end{equation*}
and the curvature-dependent function $g(s)$,
\begin{equation*}
g(s)= \varkappa(s) \left(\dfrac{\varepsilon_{+}}{\kappa_{+}^2} - \dfrac{\varepsilon_{-}}{\kappa_{-}^2}\right) \left(\dfrac{\varepsilon_{+}}{\kappa_{+}^3} + \dfrac{\varepsilon_{-}}{\kappa_{-}^3}\right)^{-1}  \: ,
\end{equation*}
is an effective geometric potential for plasmons~\cite{Longhi_2010}.
\begin{figure}[b]
\centering\includegraphics[width=0.95\linewidth]{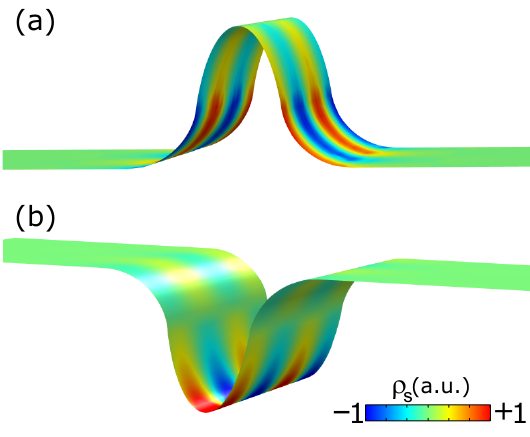}
\caption{(Color online) Channel waveguide modes found for the value of graphene conductivity $\sigma=0.00035i$ S using FEM method for $\eps_{+} = 1$, $\eps_{-} = 2$:
shown is the surface charge density distribution $\rho_s$ on the curved graphene surface. 
The frequencies of the eigenmodes plotted in (a) and (b) are $21.1$ THz and $23.9$ THz, respectively. The geometrical parameters of the curved region are: the height (depth) of the channel is 300 nm and the width of the upper (lower) elliptical segment is 200 nm, and the radius of curvature of the two lower (upper) circular segments is 140 nm, shown $y$-length is 1 $\mu$m. }
\label{fig:fig3}
\end{figure}

Remarkably, the result of Ref.~\cite{Longhi_2010} for plasmons at a curved metal-dielectric interface in the absence of graphene can be retrieved from Eq.~(\ref{eq:eqA}) by setting $\sigma(\omega)=0$, and assuming that the substrate possesses a plasma-like frequency-dependent dielectric permittivity $\eps_{-}$. In this case $ \eps_{-}/\kappa_{-} = - \eps_{+}/\kappa_{+} $, that leads to $g(s) = - R^{-1} \dfrac{\kappa_{+} \kappa_{-}}{\kappa_{+} - \kappa_{-}} = R^{-1} \dfrac{k_{\text{sp}}^2}{k_0} \left(-\dfrac{1}{\eps_{+}+\eps_{-}}\right)^{1/2}$.
Besides, a particular case characterized by a constant curvature recovers the problem of TM plasmon propagation along cylindrical graphene-coated nanowires~\cite{GrNanowire_OE_2013}.

\begin{figure}[t]
\centering\includegraphics[width=0.95\linewidth]{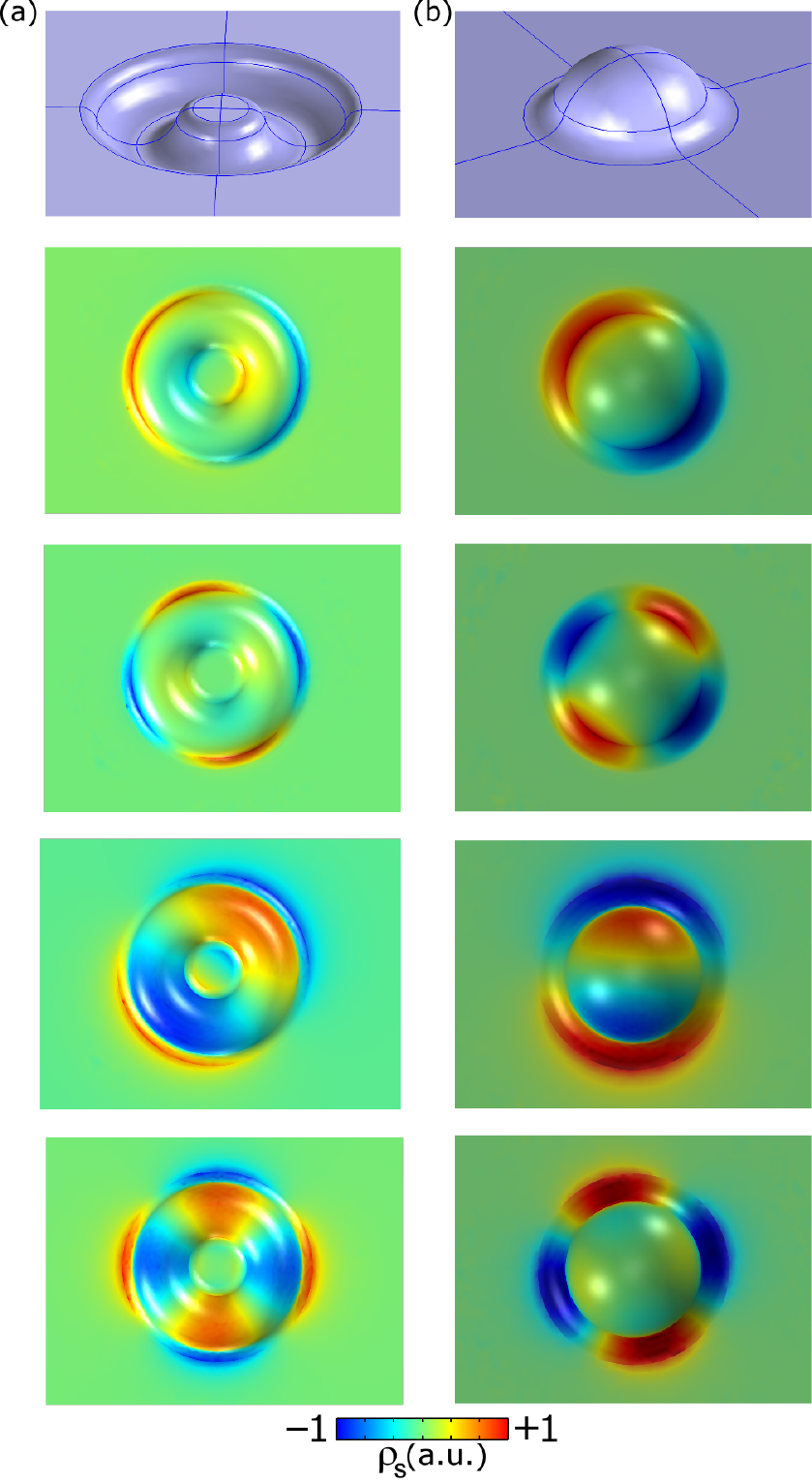}
\caption{(Color online) Geometry and localized modes of (a) circular channel (depth 100~nm, side curvature radius 33~nm) and (b) hemispherical bump (height 200~nm, side curvature radius 66~nm) found with the use of FEM method. Upper panels in (a) and (b) show profiles of the curved regions. $\eps_{+} = 1$, $\eps_{-} = 2$. Four lower panels in (a) and (b) depict 
the surface charge density distributions corresponding to the modes with increasing number of nodes in angular and radial directions. The eigenfrequencies found for the value of graphene conductivity $\sigma=0.00035i$ S (from top to bottom) (a) 11.1~THz, 14.4~THz, 15.2~THz, 21.8~THz; (b) 10.8~THz, 14.5~THz, 17.2 ~THz, 22.8~THz.}
\label{fig:fig4}
\end{figure}

It is also important to note that, as follows from Eq.~(\ref{eq:eqA}), the symmetric dielectric environment, $\eps_{-} = \eps_{+} = \eps$, suggests $g(s)=0$, and implies that no localization to the curved region is possible in the case of 1D curvature, while the loss coefficient is found to be $\gamma =\dfrac{2 \pi }{k_{\text{sp}} \omega} \sigma^{(R)} (\omega) \dfrac{\kappa^3} {\eps} $, which is in agreement with that calculated earlier in Refs.~\cite{Coupler_PRB, DissipSoliton_LPR}, where the equation for the correction $ H^{(2)}$ is written with incorporated boundary conditions using the formalism of $\delta$ functions.

As an example, we consider the interface profile of the Gaussian form. This results in inhomogeneous curvature which at some parts becomes a potential well capable of confining a mode. In our calculations, graphene is described by the surface conductivity written in terms of the Drude model $\sigma(\omega) =\displaystyle{(ie^2/\pi \hbar^2)\mathcal{E}_{F}(\omega + i\tau_{\text{intra}}^{-1})^{-1}}$, where
the Fermi energy $\mathcal{E}_{F} = 0.307$ eV, $\tau_{\text{intra}} = 0.1$ ps is the relaxation time. This case of doped graphene satisfying $\hbar\omega < \mathcal{E}_{F} $, allows neglecting the effects of interband transitions and we neglect temperature effects~\cite{sptdep}.

The potential corresponding to the Gaussian channel operates as a 2D plasmonic waveguide with trapped modes being eigenstates of the stationary Schr\"{o}dinger-like Eq.~(\ref{eq:eqA}), shown in Fig.~\ref{fig:fig2}. The first example considered here corresponds to the substrate dielectric constant exceeding that of the superstrate. Only one bound state is found for the given set of parameters with its amplitude spatial dependence $\mathcal A(z,y)=a(z)\exp{({i}\Delta k_{\text{sp}}y)}$. The position of eigenvalue $\Delta k_{\text{sp}}$ with respect to the trapping potential is indicated in Fig.~\ref{fig:fig2}(a)(top panel, red dashed line). The state appears to be less localized to the curved region with the fields primarily located near the side regions of maximal positive curvature.
The second case studied has inverted profile and also supports one localized state illustrated in Fig.~\ref{fig:fig2}(b) top panel.
One can see that for this case the field is mainly localized at the central dip (notch) having the maximal positive curvature and exhibits quite strong localization in the in-plane transverse $z$ direction.

{\em Numerical approach.} To validate the developed analytical theory, first principle numerical simulations are performed with the use of commercial full-vector FEM solver COMSOL Multiphysics. The graphene is modeled by the surface current tangential to the curved interface between the substrate and the superstrate ${\bf j}={\sigma } {\bf E}_{\tau }$. The numerical results, shown in Fig.~\ref{fig:fig2}~(a,b) bottom panel are found to be in excellent agreement with the results of our analytical theory.

While the perturbative nature of the analytical model limits its applicability to the case of small curvatures, the possibility to confine light with the curvature extends beyond this case. Moreover, one can expect that the transverse in-plane confinement of the channel modes can be further improved by increasing curvature. The particular cases of guiding by the strongly curved non-analytic graphene profiles are given in Figs.~\ref{fig:fig3}(a,b). Similar to the case of small curvatures, the mode corresponding to graphene curved towards the high index material shows more localized mode at the center of the curved region.

The proposed concept of trapping light in curved graphene can be further extended to the case of 2D profiles. Here we consider the cases of generalization of the 1D profile closed to form circular channel and hemispherical bump in 2D geometry. The resultant structures possess cylindrical symmetry and the corresponding trapped modes are expected to be quantized with respect to the azimuthal direction. Indeed, numerical results obtained in COMSOL by solving full 3D FEM problem for both cases shown in Figs.~\ref{fig:fig4}(a,b) reveal a variety of modes with increasing number of nodes as the frequency of the mode gradually increases. As opposed to the channel modes, and similar to the case of localized surface-plasmons in metallic nanoparticles, these modes are  leaky, and therefore they have a finite radiative lifetime. Nonetheless, just as in the case of plasmonic oligomers and periodically arranged metasurfaces, we expect that by building arrays of graphene bumps and dents one can engineer modes  of high quality factors limited only by internal Ohmic loss in graphene.

{\em Conclusions.} We have suggested a novel approach to confine light in curved graphene landscapes. We have shown that curved graphene allows guiding surface plasmons trapped to the curved regions. With the technique to either directly grow or transfer graphene on top of structured substrates fully developed, the demonstrated ability to confine and manipulate light with the curvature envisions photonic networks of interconnected channel waveguides, which can serve as a versatile platform for on chip graphene plasmonics integration.

{\em Acknowledgements.} This work was supported by the Australian Research Council. A part of research was carried out at the Center for Functional Nanomaterials of the Brookhaven National Laboratory supported by the U.S. Department of Energy, Office of Basic Energy Sciences under Contract No. DE-SC0012704. Z.W. acknowledges support from the Packard Fellowships for Science and Engineering and the Alfred P. Sloan Research Fellowship.

\end{document}